\documentclass[11pt]{article}
\usepackage[utf8]{inputenc}
\usepackage[T1]{fontenc}
\usepackage{lmodern}
\usepackage[a4paper,margin=1in]{geometry}
\usepackage{cite}
\usepackage{amsmath,amssymb,amsfonts}
\usepackage{algorithmic}
\usepackage{graphicx}
\usepackage{textcomp}
\usepackage{xcolor}
\usepackage{subcaption}
\usepackage{booktabs}
\usepackage{makecell}
\usepackage{newunicodechar}
\usepackage[hidelinks]{hyperref}

\newunicodechar{χ}{\ensuremath{\chi}}
\newunicodechar{ψ}{\ensuremath{\psi}}
\newunicodechar{β}{\ensuremath{\beta}}
\newunicodechar{ϕ}{\ensuremath{\phi}}

\def\BibTeX{{\rm B\kern-.05em{\sc i\kern-.025em b}\kern-.08em
    T\kern-.1667em\lower.7ex\hbox{E}\kern-.125emX}}

\title{Quantitative Assessment of Hemispheric Asymmetry in Alzheimer's Disease and Healthy Brains via QSM-Derived Biomarkers: A Proof of Concept Study}
\author{
Sadad Hasan$^{1}$, 
Prantik Paul$^{1}$, 
and Maruf Ahmed$^{1}$\\[6pt]
\small
$^{1}$Department of Electrical and Electronic Engineering\\
Bangladesh University of Engineering and Technology, Dhaka, Bangladesh\\[4pt]
\small
Corresponding author: Sadad Hasan\\
\texttt{1906163@eee.buet.ac.bd}
}

\date{}

\begin{document}

\maketitle
\begin{abstract}
Studying brain lateralization and the resulting hemispheric asymmetry have been the topic of interest for many researchers. Very few research works have focused on the differences in brain lateralization and hemispheric asymmetry caused due to neuro-degeneration. Additionally, Quantitative Susceptibility Mapping (QSM) has not been used extensively to investigate these differences. This article aims to study the differences in hemispheric asymmetry of healthy brain and Alzheimer's Disease (AD) affected brain by calculating their Myelin Hemispheric Ratio (MHR) and Oxygen Extraction Fraction Hemispheric Ratio (OEFHR) between the two hemispheres using QSM as an intermediate marker of neurodegeneration. The estimations and corresponding results show marginal yet promising distinctions between healthy subjects and AD patients. With encouraging initial results this study emphasizes the need for investigating structural brain asymmetry and symmetry as a prominent marker of AD related neurodegeneration.

\end{abstract}


\section{Introduction}

Localized neuro-degeneration correlates to structural lateralization which leads to understanding how baseline brain asymmetry influences disease vulnerability \cite{Lubben2021EnigmaImplicationsBrainAsymmetry}. Prior studies suggest that alterations in structural hemispheric asymmetry are disease-specific \cite{Mundorf2021AsymmetryCentralNervousSystem}. Some of the works have aimed to study the hemispheric asymmetry using functional MRI (fMRI) \cite{Wang2023BrainAsymmetryHemisphericNetwork} and some using diffusion MRI \cite{Chandio2026AlongTractAsymmetryAD}. The goal of this study is to test the potential biomarkers of hemispheric asymmetry alterations occurring at neuro-degeneration caused by Alzheimer's Disease (AD). Though β-amyloid deposition and accumulation differences are hallmarks of AD \cite{Azargoonjahromi2024DualityAmyloidBeta, Hampel2021AmyloidBetaPathway}, more biomarkers can potentially lead to understanding of the disease pathology and prognosis. This will eventually lead to delayed neuro-degeneration and possible remission. That is why, Myelin Hemispheric Ratio (MHR) and Oxygen Extraction Fraction Hemispheric Ratio (OEFHR) have been proposed and calculated through this work so that the differences in hemispheric asymmetry between healthy brains and AD-affected brains can be identified and possibly quantified. The paper aims to open the possible doors and widen the horizons for studying hemispheric asymmetry and the possible alterations occurring due to different neuro-degenerative diseases by re-utilizing existing biomarkers for new and different diagnostic scenarios. Multi-Echo Gradient Recalled Echo (MEGRE) Magnetic Resonance (MR) images were utilized to generate Quantitative Susceptibility Mapping (QSM) based on which we calculated the Oxygen Extraction Fraction (OEF) and myelin content followed by the OEF and myelin hemispheric ratios for both healthy controls and AD patients. In this paper, a comparison between STAR-QSM and DECOMPOSE-QSM based MHR and OEHR calculations has been done.  The aim of this study is to provide a line of thought which will help subsequent researchers to quantify hemispheric asymmetry and potentially use it to test cognition and neurodegeneration. Besides, the study also highlights the significance of pre-processing or selection of QSM methodology in this case, on the estimation of quantitative markers.

\section{Literature review}

QSM is emerging as a useful tool for disease diagnosis due to its ability to provide excellent contrast of iron-rich deep nuclei and white matter bundles from surrounding tissues \cite{Li2011}. Several studies have explored the potential of QSM in clinical applications \cite{Liu2015QSMContrast, Harada2022, Ravanfar2021, Luyken2025}. Studies have proven that QSM can be an effective tool to estimate the iron content of human brain \cite{Ravanfar2021, Luyken2025, Langkammer2012}. Langkammer et al. (2012) first showed that QSM can be used to efficiently estimate iron concentration in brain \cite{Langkammer2012}. Kumar et al. (2021) aimed to highlight and segment paramagnetic susceptibility sources, mainly iron and diamagnetic susceptibility sources, mainly myelin and Calcium at ultra-high value of magnetic field \cite{Kumar2021QSM94T}. Likewise, QSM can be used to quantify certain biomarkers which otherwise are measured by other methods. However, conventional bulk QSM can not disentangle the reconstructed bulk susceptibility sources within a voxel \cite{Hametner2018}. The cancellation effects occur as the bulk susceptibility is the summation of paramagnetic and diamagnetic contributions. Shin et al. (2021) developed a bio-physical model which is used to implement $\chi$-separation, a method to generate the voxel-wise distributions of the two sources of susceptibility \cite{Shin2021ChiSeparation}. In the same year, DECOMPOSE-QSM algorithm was introduced which decomposes QSM into sub-voxel paramagnetic and diamagnetic components \cite{DECOMPOSEQSM}. Source-separation methodologies have developed and diversified a lot. Solutions like APART-QSM and QSM-ARCS aim to improve susceptibility source separation \cite{Li2023APARTQSM, Kan2024QSMARCS}. Several studies establish the fact that susceptibility source separation can help comprehend myelin-related changes and conditions which is typically not possible using conventional bulk QSM \cite{AHMED2023120357,Zhu2024MRSusceptibilitySeparationRRMS, Muller2024RemyelinationChiSeparation}. Thus, to estimate and calculate susceptibility-based biomarkers with a higher accuracy, susceptibility source separation based QSM is a better choice compared to bulk QSM.

The idea of estimating myelin to study AD had been executed in several research works \cite{Bartzokis2004, Bartzokis2011, Nasrabady2018, Mayo2016}. The works of Bartzokis in 2004 and 2011 led to the establishment of the fact that myelin breakdown  progressively occurs in AD patients \cite{Bartzokis2004, Bartzokis2011}. Nasrabady et al. (2018) showed that white matter changes including myelin alterations are prominent features of AD \cite{Nasrabady2018}. Besides, Mayo et al. (2016) noticed significant differences in Fractional Anisotropy (FA) and Mean Diffusivity (MD) of AD patients and those of healthy individuals which suggests microstructural changes in white matter due to AD \cite{Mayo2016}. Many non-MRI methods, mainly PET imaging, have been used to estimate myelin \cite{Moscoso2022, Zeydan2019, Xu2025,vanderWeijden2022} But MRI-based methods, particularly, QSM-based susceptibility separation methods are the most promising \cite{vanderWeijden2023}. Many of the contemporary research works consist of MRI-based myelin estimation. Myelin-Water Fraction (MWF), a well-known myelin-health quantifier is used in many studies \cite{Lee2021MyelinWaterGuide, Kim2023MRFingerprintingMWF, Puliyakkara2026MWISystematicReview, Lee2024MyelinWaterTranscriptome}. Mangeat et al. (2015) combined magnetisation transfer, $T_2^\ast$, and $B_0$-orientation to characterize myelo-architecture \cite{Mangeat2015}. Mancini et al. (2018) used $g$-ratio, which is a myelination measure \cite{Mancini2018GRatioConnectomics}. Sui et al. (2025) used magnetisation-transfer and $g$-ratio in their study \cite{Sui2025GRatioPsychosis}. Chopra et al. (2018) used T1w/T2w technique to estimate brain myelin content \cite{Chopra2018MyelinatedTractsProcessingSpeed}. Shafee et al. (2015) estimated myelination using partial-volume corrected MRI \cite{Shafee2015GrayMatterMyelination}. Susceptibility-based myelin imaging is one of the most widely accepted and used methods in recent MRI-based studies. A considerable number of research works have isolated the diamagnetic component which mainly comes from myelin \cite{DECOMPOSEQSM, Shin2021ChiSeparation, Li2023APARTQSM, Kan2024QSMARCS}. Ahmed et al. (2023) used the diamagnetic component from DECOMPOSE-QSM to study AD-related neurodegenerative change \cite{AHMED2023120357}.

Regarding the study of human brain asymmetry, a review of this has been done by Toga et al. (2003) \cite{Toga2003}. Besides, the existence of hemispheric asymmetry of human brain has been studied and eventually established in existing literature \cite{Wan2024MicrostructuralAsymmetry, Korbmacher2024BrainAsymmetries }. This lateralization is present from childhood and then continues to adulthood \cite{Parekh2023WhiteMatterLateralizationSPD, Kanakaraj2025LifespanAsymmetryTracts}. It is a proven fact by research that hemispheric asymmetry reduces in human brain due to aging \cite{Cabeza2002, Dolcos2002, Park2009, Korbmacher2024BrainAsymmetries}. Timmers et al. (2016) studied white matter abnormalities between healthy controls and patients with classic galactosemia \cite{Timmers2016DTINODDI}. Even between patients of two different diseases, that is, AD and Mild Cognitive Impairment (MCI), difference in asymmetry may exist \cite{Yang2017}. Similarly, healthy controls and AD patients have differences in the white-matter asymmetry \cite{Chandio2026AlongTractAsymmetryAD}. The age-related asymmetry reduction and asymmetry differences due to neurological disorders are the main driving factors for conducting a hemispheric study of OEF and myelin content and thus contrast their differences.

Oxygen extraction fraction (OEF), the proportion of oxygen removed from arterial blood during capillary transit, is a direct index of the balance between cerebral oxygen delivery and demand \cite{Jiang2022}. Because deoxyhaemoglobin is paramagnetic while oxyhaemoglobin is weakly diamagnetic, venous blood carries a susceptibility offset relative to parenchyma in proportion to its deoxyhaemoglobin content, so venous oxygen saturation can be recovered directly from QSM without radiotracers or gas challenges \cite{Kudo2016, Jiang2022}. Kudo et al.\ (2016) validated this principle against the ${}^{15}$O-PET reference standard in 26 patients with unilateral steno-occlusive disease and 15 normal subjects, reporting a significant correlation of the inter-hemispheric OEF ratio ($r = 0.62$, $p < 0.001$) with sensitivity 0.63 and specificity 0.89 for detecting elevated OEF \cite{Kudo2016}; the sliding-window volume-of-interest (VOI) formulation introduced there is the framework adopted here. Fan et al.\ (2020) subsequently confirmed the expected elevation of QSM-derived OEF in the affected hemisphere in acute ischaemic stroke \cite{Fan2020}. Two reconstruction-side factors limit the measurement. First, veins whose diameter spans only $1.5$--$4$ voxels suffer partial-volume dilution of their apparent susceptibility, biasing OEF downwards \cite{Ward2017}. Second, the ill-posed dipole inversion produces streaking artifact near high-dynamic-range sources such as veins, contaminating exactly the voxels the estimate depends on; STAR-QSM suppresses this through two-level regularisation while preserving sharp venous structure \cite{Wei2015}, which motivates its use as one of the two susceptibility inputs in this work.

A less-studied limitation is that bulk susceptibility in any voxel is the algebraic sum of paramagnetic and diamagnetic contributions, so co-localised myelin can partially cancel the deoxyhaemoglobin signal on which OEF depends \cite{Hametner2018}. DECOMPOSE-QSM resolves this by fitting a three-pool complex GRE signal model that separates each voxel into paramagnetic component susceptibility (PCS) and diamagnetic component susceptibility (DCS) \cite{DECOMPOSEQSM}. Since venous contrast arises exclusively from paramagnetic deoxyhaemoglobin, PCS is in principle a more specific input for OEF than total susceptibility; recent work supports this claim in case of cerebral small-vessel disease \cite{Zhou2026ParamagneticSusceptibility} and in neonatal cerebral oximetry, where paramagnetic-component saturations were closer to reference values although more variable \cite{Carmichael2025SourceSeparationOxygenation}. The expected direction of change in AD is not, however, that of cerebrovascular disease. Jiang et al.\ (2020) showed that OEF is differentially altered by the two pathologies: reduced in Alzheimer's disease, where diminished neural activity lowers oxygen demand, but elevated where vascular pathology restricts supply \cite{Jiang2020ADVascular}. This is consistent with the reduced global oxidative metabolism reported in amnestic MCI \cite{Thomas2017MCIMetabolism} and with the downward OEF trend observed across the AD clinical spectrum \cite{Hosseini2023QSMOEFADSpectrum}, although reports of increased venous susceptibility in AD \cite{Liu2021ADVenousOxygen} show that the literature is not unanimous and that the direction is cohort-dependent. Since hemispheric metabolic asymmetry is present in the healthy brain and is altered by neurodegeneration \cite{Toga2003, Yang2017, Chandio2026AlongTractAsymmetryAD}, we evaluate lateralised OEF, that is, OEFHR, and do so under two susceptibility reconstruction approaches rather than one.

\section{Methodology}

\subsection{Dataset Used}
We used $T_2^{*}$ weighted MR images of 10 healthy individuals and 10 AD patients. The dataset consisted of both magnitude and phase images. The dataset was procured by University of California (UC) San Francisco and curated by Professor Chunlei Liu of University of California (UC) Berkeley. 3 Tesla Siemens MRI machine with 32-channel phased-array  coil was used for the MEGRE scans. The echo time (TE) was 14 milliseconds  and repetition time (TR) was 20 milliseconds with an echo count of 8. 

\begin{figure}[htbp]
    \centering

    \begin{subfigure}[t]{0.45\columnwidth}
        \centering
        \includegraphics[
            width=0.7\linewidth,
            height=4.8cm
        ]{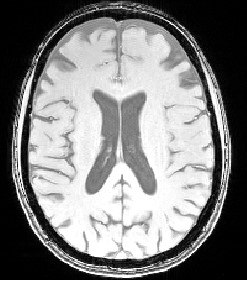}
        \caption{Magnitude}
        \label{fig:magnitude}
    \end{subfigure}
    \hfill
    \begin{subfigure}[t]{0.45\columnwidth}
        \centering
        \includegraphics[
            width=0.7\linewidth,
            height=4.8cm
        ]{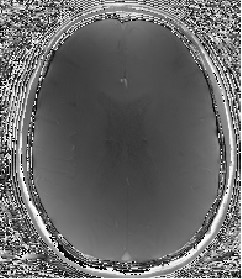}
        \caption{Phase}
        \label{fig:phase}
    \end{subfigure}

    \caption{Raw magnitude and phase data from the first echo of a patient with AD.}
    \label{fig:ds}
\end{figure}

\subsection{Generation of STAR-QSM images}\label{sec3}
Quantitative Susceptibility Mapping (QSM) is an advanced MRI technique which estimates
the magnetic susceptibility distribution in tissues involving multiple stages, from MRI data
acquisition to final susceptibility map reconstruction. The STAR-QSM, that is, STreaking Artifact Reduction for QSM algorithm is designed to minimize streaking artifacts while preserving fine anatomical details \cite{Wei2015}. The key features of the algorithm include:

\begin{itemize}
    \item A two-level regularization approach for handling large and small susceptibility sources separately.
    \item Dynamic adjustment of the regularization parameter to ensure a balance between artifact suppression and the preservation of fine anatomical features.
    \item The ability to process sources with large dynamic ranges, such as hemorrhages, calcification, and gadolinium contrast agents.
\end{itemize}

\begin{figure}[htbp]
    \centering
    \includegraphics[width=\textwidth]{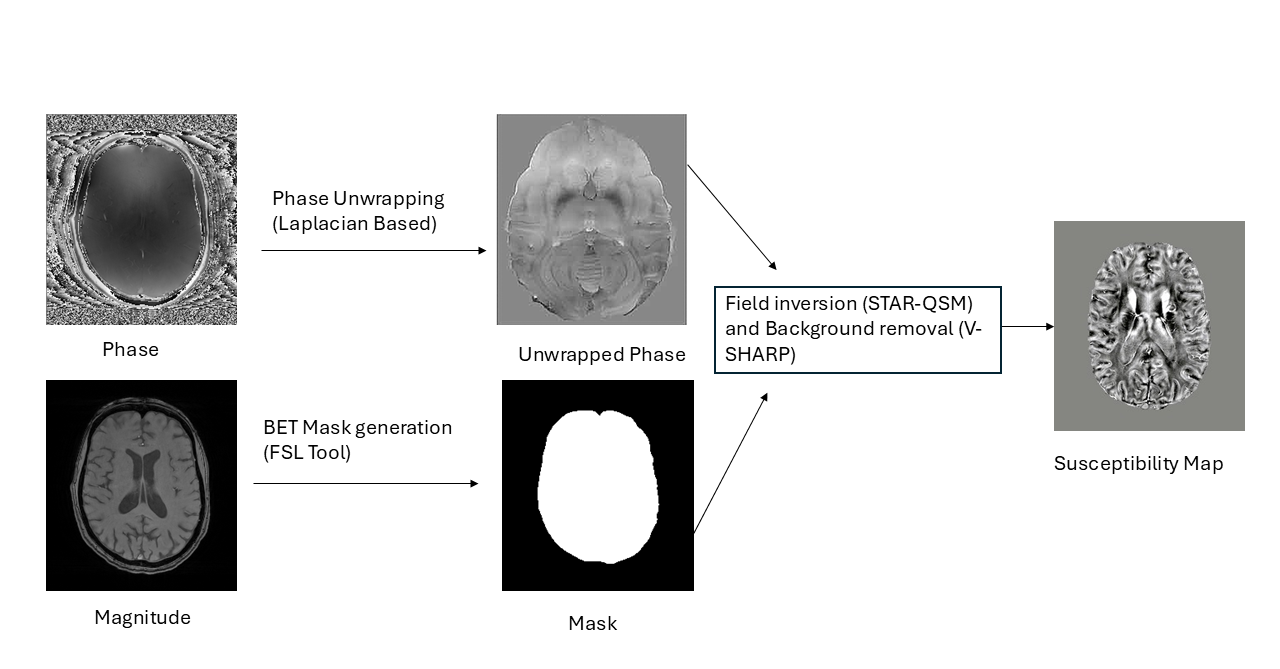}
    \caption{STAR-QSM Generation Pipeline.}
    \label{starqsmgen}
\end{figure}

STAR QSM reconstructs tissue magnetic susceptibility (χ) from the phase evolution of gradient-echo MRI, where the measured phase (ψ) is related to χ through a dipole convolution model in Fourier space, i.e., \begin{equation}
\psi = \mathcal{F}^{-1} \left( D_2 \cdot \mathcal{F} \chi \right)
\label{eq:phase_sus}
\end{equation} 
, where $D_2$ denotes the discrete magnetic dipole convolution kernel computed in \textit{k}-space and $F$ denotes Fourier transform.
STAR-QSM generation methodology has been described in detail by Wei et al. (2015) \cite{Wei2015}. However, for the ease of understanding, the steps are discussed in this paper as well.
Fig.~\ref{starqsmgen} demonstrates the pipeline for STAR-QSM generation.

\subsubsection{Preprocessing and field preparation}
After converting the scanner DICOM data into NIfTI volumes, producing multi-echo magnitude and phase stacks along with the acquisition parameters, magnitude and phase images were combined to form complex GRE data as 
\begin{equation}
S = M \cdot e^{i\phi}
\end{equation}
, where M is magnitude and ϕ is the wrapped phase.
 For handling multi-echo phase prior to inversion, several approaches can be taken and of them a common and widely accepted approach is to estimate a frequency map by fitting phase or the complex signal across echo times. This can improve SNR and reduce echo-to-echo inconsistencies. Alternatively, each echo can be processed independently and combined later using magnitude-weighted averaging or complex fitting. In our implementation, we took the latter approach and so the processing was performed per time-point/echo as described in the STAR-QSM generation algorithm. That is, we processed signal of each echo-time individually and thus enabled consistent masking, unwrapping, and background field correction across the dataset.

 In order to reduce or effectively remove boundary-related artifacts and thus improve the numerical stability, FFT-based zero-padding was applied in \textit{k}-space for each time-point by padding 64 voxels along the \textit{z}-direction and then it was transformed back to image space. 

\subsubsection{Brain masking}

There are many software tools to generate a brain mask, such as, FSL, We used the BET ( Brain Extraction Tool) of FSL to generate the brain mask and saved for subsequent usage \cite{Jenkinson2012}.

\subsubsection{Phase unwrapping}

Since the measured phase is wrapped to [-$\pi$,$\pi$], phase unwrapping is required prior to local field estimation. Multiple valid unwrapping approaches exist, including Laplacian-based unwrapping, path-following/region-growing methods, and quality-guided techniques. For our task, we performed Laplacian-based phase unwrapping.

\subsubsection{Background field removal}

The unwrapped phase contains both local tissue contributions and background fields arising from sources lying outside the region of interest. The background removal can be performed using several well-established techniques, such as SHARP/V-SHARP, RESHARP, PDF, or LBV, each with different trade-offs regarding mask erosion, robustness near boundaries, and sensitivity to noise. We chose V-SHARP due to its effectiveness and widespread use in QSM generation pipelines.

\subsubsection{STAR-QSM inversion}

After normalizing the phase, a two-level regularization framework separately reconstructs strong and weak susceptibility sources. Using the relationship between phase and magnetic susceptibility, we set the regularization parameter such that it can estimate the susceptibility from strong sources using the equation:
\begin{equation}
\chi = \min_{\chi} \left( \| \mathcal{F}^{-1}(D_2 \cdot \mathcal{F}(\chi)) - \psi \|_2 + \lambda \| W \cdot G \cdot \chi \|_1 \right)
\label{suscalc}
\end{equation}
, where $\lambda$ is the regularization parameter, $W$ are the weighting factors derived from initial susceptibility estimates, $G$ is  Gradient operator, $D_2$ is the discrete magnetic dipole convolution kernel computed in \textit{k}-space, $\psi$ is the normalized phase, and $F$ indicates Fourier Transform.

Then using the Equation \eqref{eq:phase_sus} and getting the phase component from strong sources, the phase component of tissues using the equation: 
\begin{equation}
    \psi_{tissue} = \psi_{local} - \psi_{strong}.
\end{equation}
Using a small regularization parameter $\beta$, instead of $\lambda$ in Equation \eqref{suscalc}, we get the tissue susceptibility. Combining the tissue susceptibility and and susceptibility from strong sources, we get the final susceptibility maps \cite{Wei2015}. To generate STAR-QSM, we used STISuite which is a MATLAB Toolbox developed and maintained by the research group of Professor Chunlei Liu at UC Berkeley \cite{STISuite}. The images are displayed using ITK-SNAP \cite{Yushkevich2006}.

\begin{figure}[htbp]
    \centering
    \includegraphics[width=0.49\columnwidth]{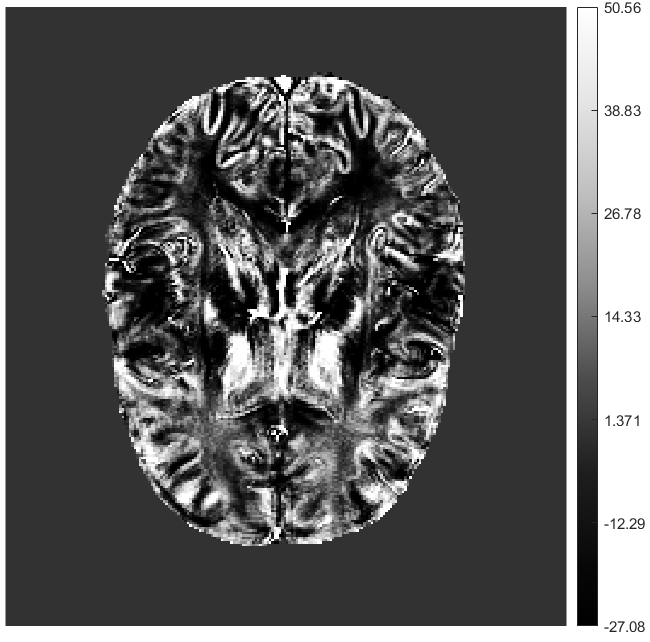}
    \caption{Susceptibility map (in ppb) generated for a brain affected by AD using STAR-QSM.}
    \label{star-qsm_sample}
\end{figure}

\subsection{Generation of DECOMPOSE-QSM images}
\begin{figure}[htbp]
    \centering
    \includegraphics[width=\textwidth]{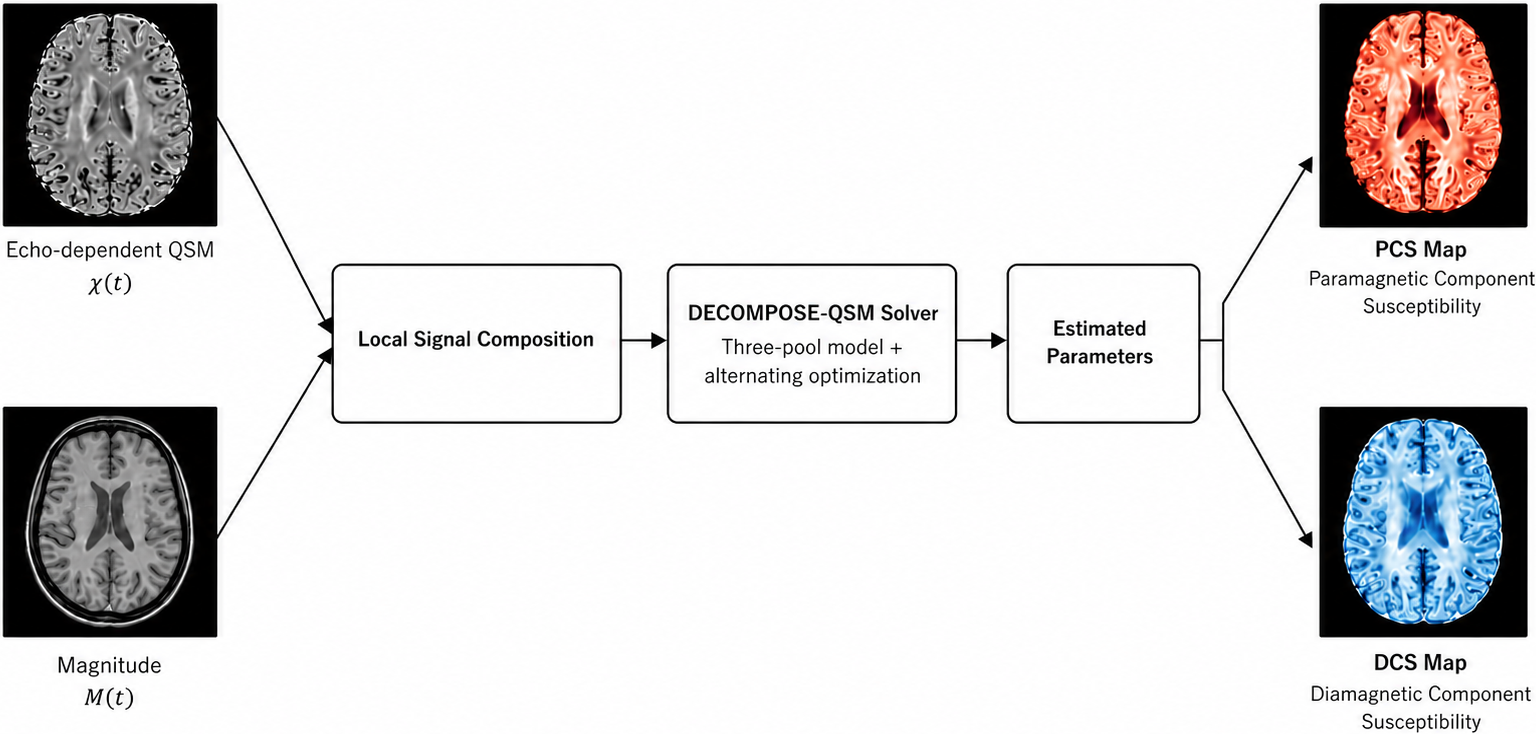}
    \caption{DECOMPOSE-QSM Generation Pipeline.}
    \label{decomposeqsmgen}
\end{figure}

DECOMPOSE-QSM, that is, DiamagnEtic COMponent and Paramagnetic cOmponent SEparation of QSM, is a method which decomposes the conventional QSM into sub-voxel level paramagnetic and diamagnetic components \cite{DECOMPOSEQSM}. It fits a three-pool complex GRE signal model to estimate paramagnetic component susceptibility (PCS) and diamagnetic component susceptibility (DCS). For generating DECOMPOSE-QSM, we used the codes developed by the research group of Professor
Chunlei Liu at UC Berkeley. Even though Chen et al. (2021) explained DECOMPOSE-QSM generation process in detail \cite{DECOMPOSEQSM}, for the ease of understanding, it has been described in detail. Fig.~\ref{decomposeqsmgen} demonstrates the pipeline for DECOMPOSE-QSM generation.

\subsubsection{QSM preprocessing}

 Computing the STAR-QSM for each echo time is a necessity to fully capture the signal behavior in voxels containing mixtures of diamagnetic and paramagnetic sources. These echo-dependent susceptibility maps form the basis for synthesizing the local signal used in DECOMPOSE-QSM.

\subsubsection{Synthesis of the local complex signal \& Three-pool sub-voxel signal model}
 Instead of directly fitting the raw complex GRE signal, DECOMPOSE-QSM synthesizes a local complex signal using the measured magnitude and the phase predicted from the echo-dependent QSM to ensure that phase contributions from sources outside the voxel are removed and that the observed signal is consistent with the voxel-scale susceptibility distribution. Each voxel is modeled as a mixture of three sub-voxel signal pools, Paramagnetic component with susceptibility $\chi_+$ and signal fraction $C_+$, Diamagnetic component with susceptibility $\chi_-$ and signal fraction and $C_-$, Neutral component with zero susceptibility and signal fraction $C_0$ where sum of the fractions equals 1. Under the static dephasing regime, the transverse relaxation rate of the susceptibility components is assumed to be linearly related to susceptibility magnitude:
\begin{equation}
R_{2,+,-}^* = a|\chi_{+,-}| + R_{2,0}^*,
\label{eq:r2star_model}
\end{equation}
, where $a$ is a proportionality constant and $R_{2,0}^*$ denotes the baseline relaxation rate common to all components.

The total GRE signal of a voxel is modeled as a weighted sum of three complex exponential terms:
\begin{equation}
\begin{aligned}
S(t) ={}&
C_+ \exp\left[-\left(a\chi_+ + R_{2,0}^* + i\frac{2}{3}\gamma B_0\chi_+\right)t\right] \\
&+ C_- \exp\left[-\left(-a\chi_- + R_{2,0}^* + i\frac{2}{3}\gamma B_0\chi_-\right)t\right] \\
&+ C_0 \exp(-R_{2,0}^*t).
\end{aligned}
\label{eq:signal_model}
\end{equation}

\subsubsection{Inverse problem formulation and optimization}
The unknown parameter set
\begin{equation}
\Theta = \{C_+, C_-, C_0, \chi_+, \chi_-, R_{2,0}^*\}
\end{equation}
is estimated by minimizing the discrepancy between the modeled signal and the synthesized local signal:
\begin{equation}
\min_{\Theta} \sum_t \left\| S(t;\Theta) - y(t) \right\|_2^2.
\label{eq:objective_function}
\end{equation}

This nonlinear inverse problem is solved using an alternating-direction optimization strategy, in which subsets of parameters are updated sequentially while holding the remaining parameters fixed.

With the susceptibilities $\chi_+$, $\chi_-$ and the baseline relaxation rate $R_{2,0}^*$ fixed, the signal fractions $C_+$, $C_-$, and $C_0$ are estimated via a constrained least-squares problem:
\begin{equation}
\min_{C_+,C_-,C_0} \sum_t \left\| S(t) - y(t) \right\|_2^2
\quad
\text{s.t. } C_i \ge 0,\; \sum_i C_i = 1.
\label{eq:concentration_estimation}
\end{equation}

Next, with $C_+$, $C_-$, $C_0$, $\chi_+$, and $\chi_-$ held fixed, the baseline transverse relaxation rate $R_{2,0}^*$ is updated in the logarithmic domain:
\begin{equation}
\min_{R_{2,0}^*} \sum_t
\left\| \log S(t;R_{2,0}^*) - \log y(t) \right\|_2^2,
\quad R_{2,0}^* > 0.
\label{eq:r2star_estimation}
\end{equation}

Finally, the susceptibilities $\chi_+$ and $\chi_-$ are estimated by minimizing the log-domain residual while enforcing physical bounds:
\begin{equation}
\min_{\chi_+,\chi_-} \sum_t
\left\| \log S(t;\chi_+,\chi_-) - \log y(t) \right\|_2^2,
\quad 0 < |\chi_{\pm}| < \chi_{\max}.
\label{eq:susceptibility_estimation}
\end{equation}

These parameter updates are performed sequentially and repeated iteratively until convergence.

\subsubsection{Construction of PCS and DCS maps}
The PCS and DCS are defined as effective susceptibility values under the hypothetical condition that only one susceptibility source exists alongside the neutral medium:
{\footnotesize
\begin{equation}
PCS =
\frac{
\sum_t \angle\left(
C_+ e^{-(a\chi_+ + R_{2,0}^* + i\frac{2}{3}\gamma B_0\chi_+)t}
+
(C_0+C_-)e^{-R_{2,0}^*t}
\right)
}{
\frac{2}{3}\gamma B_0 \sum_t t
},
\label{eq:PCS}
\end{equation}

\begin{equation}
DCS =
\frac{
\sum_t \angle\left(
C_- e^{-(-a\chi_- + R_{2,0}^* + i\frac{2}{3}\gamma B_0\chi_-)t}
+
(C_0+C_+)e^{-R_{2,0}^*t}
\right)
}{
\frac{2}{3}\gamma B_0 \sum_t t
}.
\label{eq:DCS}
\end{equation}
}

\begin{figure}[htbp]
    \centering

    \begin{subfigure}{0.49\columnwidth}
        \centering
        \includegraphics[width=\linewidth]{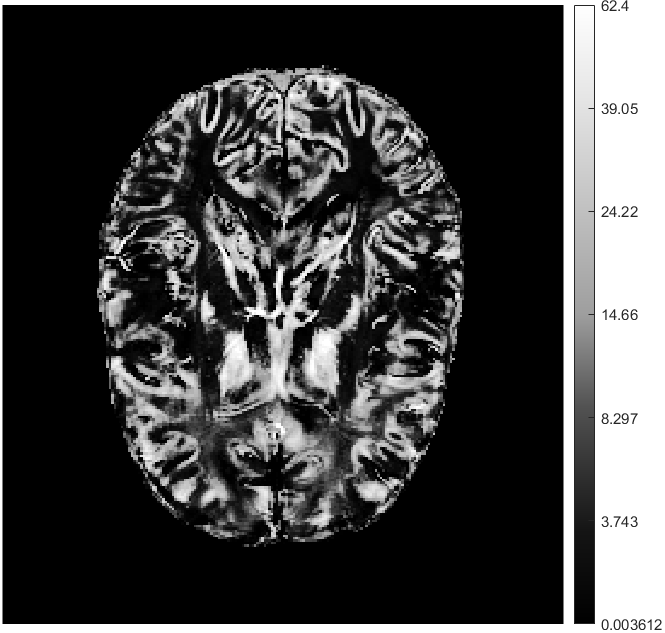}
        \caption{PCS Map}
        \label{pcs}
    \end{subfigure}
    \hfill
    \begin{subfigure}{0.49\columnwidth}
        \centering
        \includegraphics[width=\linewidth]{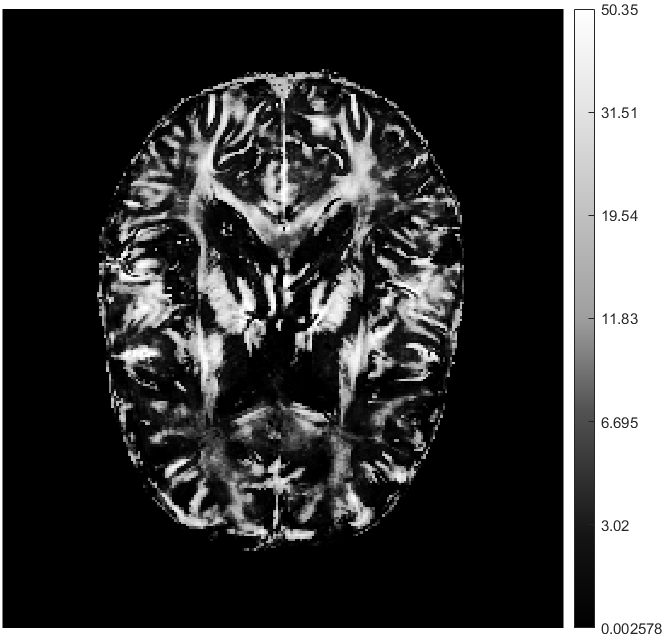}
        \caption{|DCS| Map}
        \label{dcs}
    \end{subfigure}

    \caption{ PCS and |DCS| maps (in ppb) generated using DECOMPOSE-QSM for a brain affected by AD.}
    \label{fig:hr_permutation_test}
\end{figure}

\subsection{Calculation of Myelin Content}
Myelin content is one of the main contributors of the diamagnetic component of susceptibility \cite{AHMED2023120357, Lee2024}. For STAR-QSM based calculation of myelin content, only the voxels which show diamagnetic susceptibility are taken and the summation of their values can be divided by the total number of voxels of the region or hemisphere being calculated. For DECOMPOSE-QSM based calculation of myelin content, the DCS map was used where the summation of diamagnetic susceptibility of all voxels was divided by the number of voxels of a specific region or hemisphere to estimate the myelin content. Since our goal is to study the level of hemispheric asymmetry, we calculated the Myelin Hemispheric Ratio (MHR) which was obtained by dividing the myelin average of the right hemisphere of a specific subject by the myelin average of the left hemisphere of the corresponding subject. 

\subsection{OEF Estimation from STAR-QSM and DECOMPOSE-QSM}
\label{sec:oef}

OEF was estimated twice per subject from two susceptibility inputs: the total STAR-QSM map $\chi_{\mathrm{STAR}}$, and the paramagnetic component susceptibility (PCS) map returned by DECOMPOSE-QSM. Because DECOMPOSE-QSM takes the STAR-QSM reconstruction as its input and resolves it into sub-voxel components \cite{DECOMPOSEQSM}, the two pipelines are identical in every processing step except the susceptibility image passed to the estimator. Any difference in the resulting OEF is therefore attributable to susceptibility source composition rather than to masking, thresholding or spatial averaging. The DCS map, which reflects diamagnetic sources such as myelin \cite{AHMED2023120357, Lee2024}, is not used in the OEF pipeline. This is because as haemoglobin, which is used to estimate OEF is paramagnetic in nature.

\subsubsection{Venous Voxel Selection}
\label{subsec:oef_vein}

Brain extraction used FSL BET \cite{Jenkinson2012}; the resulting mask was retained for all subsequent steps, and non-finite or zero-valued voxels inside it were discarded. Following the sliding-window VOI formulation of Kudo et al.\ \cite{Kudo2016}, a VOI of $100 \times 100 \times 25$ voxels was translated across the volume in steps of $25 \times 25 \times 6$ voxels. Within each VOI, voxels exceeding the local mean by two local standard deviations were labelled venous,
\begin{equation}
V(\mathbf{r}) =
\begin{cases}
1, & \chi(\mathbf{r}) \geq \mu_{\mathrm{VOI}} + 2\sigma_{\mathrm{VOI}}\\
0, & \text{otherwise}
\end{cases}
\label{eq:vein}
\end{equation}
, where $\chi$ denotes $\chi_{\mathrm{STAR}}$ or PCS according to the pipeline. Because the criterion is referenced to the local mean and standard deviation, it adapts automatically to the differing absolute scales of the two maps and requires no map-specific tuning. The venous--tissue susceptibility difference for that VOI is
\begin{equation}
\Delta\chi = \overline{\chi}_{\mathrm{vein}} - \overline{\chi}_{\mathrm{tissue}},
\label{eq:dchi}
\end{equation}
with $\overline{\chi}_{\mathrm{tissue}}$ the mean over the remaining valid voxels of the same VOI.

\subsubsection{OEF Model}
\label{subsec:oef_model}

Venous susceptibility relative to tissue scales with deoxyhaemoglobin concentration as $\Delta\chi = \Delta\chi_{do} \cdot \mathit{Hct} \cdot (1 - Y_v)$, where $Y_v$ is venous oxygen saturation \cite{Spees2001, Ward2017}. Taking arterial saturation as $Y_a \approx 1$ gives $\mathrm{OEF} \approx 1 - Y_v$, so that
\begin{equation}
\mathrm{OEF} = \frac{\Delta\chi \cdot P_v}{\Delta\chi_{do} \cdot \mathit{Hct}}.
\label{eq:oef}
\end{equation}
We used $\Delta\chi_{do} = 0.27$~ppm in CGS units, equivalently $4\pi \times 0.27$~ppm in the SI convention in which the susceptibility maps are expressed \cite{Spees2001}; a fixed $\mathit{Hct} = 0.40$; and the partial-volume correction factor $P_v = 1.39$ as inspired from the works of Kudo et al. (2016) \cite{Kudo2016} to compensate the underestimation of venous susceptibility in vessels only a few voxels wide \cite{Ward2017}. All three constants were held fixed across subjects and pipelines, so that group and hemispheric contrasts are not driven by subject-specific parameter choices. VOI positions returning non-finite values, or values outside the physiological interval $(0,1)$, were discarded.

\subsubsection{Hemispheric Aggregation}
\label{subsec:oef_hemi}

Volumes were aligned so that the mid-sagittal plane coincided with the geometric centre of the left--right image axis, and each accepted VOI was assigned to the left or right hemisphere according to the position of its centre relative to the midline coordinate $x_{\mathrm{mid}}$. Hemispheric OEF was taken as the mean over accepted VOI positions within each half, yielding four values per subject: $\mathrm{OEF}_{L}$ and $\mathrm{OEF}_{R}$ for each of the two pipelines. We finally calculated the OEF Hemispheric Ratio (OEFHR) by dividing the OEF average of the right hemisphere of a specific subject by the OEF average of the left hemisphere of the corresponding subject for both STAR-QSM and DECOMPOSE-QSM based calculations.

\section{Results \& Discussion}

Healthy controls showed stronger asymmetry than AD which can be comprehended from the higher Myelin Hemispheric Ratio values of healthy controls. Table~\ref{tab:hr_random_subset_stats} consists of the MHR-based comparison between Healthy and AD groups separately for DECOMPOSE-QSM and STAR-QSM. The difference of Myelin Hemispheric Ratio between the two groups for two different QSM methods is not statistically significant though healthy subjects showed higher mean MHR than AD. Table~\ref{tab:hr_mixed_anova_random_subset} proves that different QSM methods produced different MHR values and the Healthy-AD difference also varied across the two methods. The results suggest that STAR-QSM produces higher MHR than DECOMPOSE in both groups and healthy groups show a larger increase in MHR from DECOMPOSE-QSM to STAR-QSM than the AD group. Table~\ref{tab:oefhr_stats} shows that the difference in Oxygen Extraction Fraction Hemispheric Ratio (OEFHR) between the two groups is not statistically significant even though healthy subjects possess higher mean OEFHR than AD. Table~\ref{tab:oefhr_mixed_anova} proves that difference in QSM methods result in different OEFHR values.

\begin{figure}[htbp]
    \centering

    \begin{subfigure}{0.49\columnwidth}
        \centering
        \includegraphics[width=\linewidth]{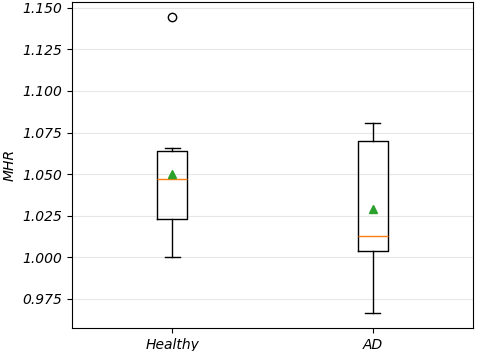}
        \caption{MHR value computed using DECOMPOSE-QSM}
        \label{fig:decompose_mhr_box}
    \end{subfigure}
    \hfill
    \begin{subfigure}{0.49\columnwidth}
        \centering
        \includegraphics[width=\linewidth]{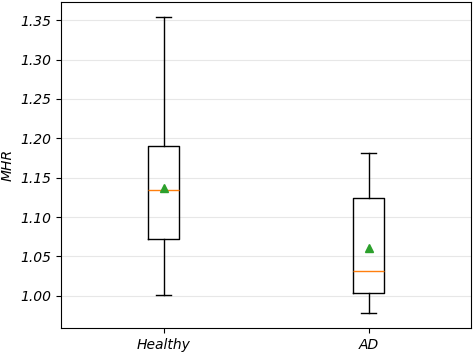}
        \caption{MHR value computed using STAR-QSM}
        \label{fig:star_mhr_box}
    \end{subfigure}

    \caption{Box Plot of MHR values of Healthy and AD groups using DECOMPOSE-QSM and STAR-QSM .}
    \label{fig:mhr_boxplots}
\end{figure}

\begin{figure}[htbp]
    \centering

    \begin{subfigure}{0.49\columnwidth}
        \centering
        \includegraphics[width=\linewidth]{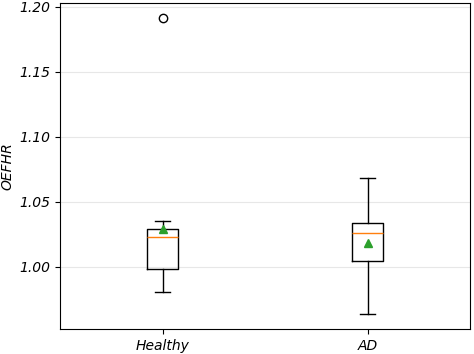}
        \caption{OEFHR value computed using DECOMPOSE-QSM}
        \label{fig:decompose_oefhr_box}
    \end{subfigure}
    \hfill
    \begin{subfigure}{0.49\columnwidth}
        \centering
        \includegraphics[width=\linewidth]{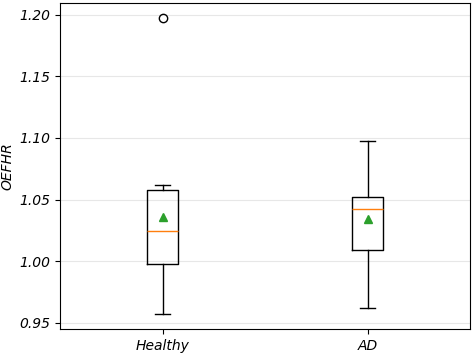}
        \caption{OEFHR value computed using STAR-QSM}
        \label{fig:star_oefhr_box}
    \end{subfigure}

    \caption{Box Plot of OEFHR values of Healthy and AD groups using DECOMPOSE-QSM and STAR-QSM .}
    \label{fig:oefhr_boxplots}
\end{figure}


\begin{figure}[htbp]
    \centering

    \begin{subfigure}{0.49\columnwidth}
        \centering
        \includegraphics[width=\linewidth]{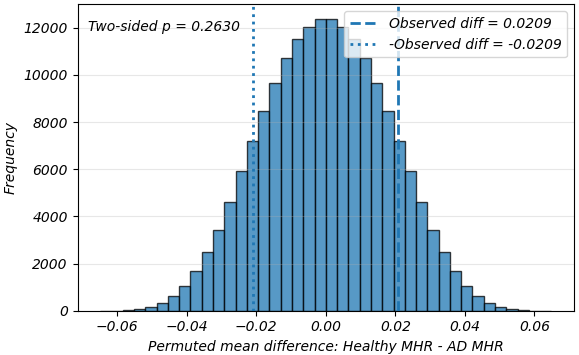}
        \caption{Permutation of MHR calculated using DECOMPOSE-QSM}
        \label{fig:decompose_permutation_mhr}
    \end{subfigure}
    \hfill
    \begin{subfigure}{0.49\columnwidth}
        \centering
        \includegraphics[width=\linewidth]{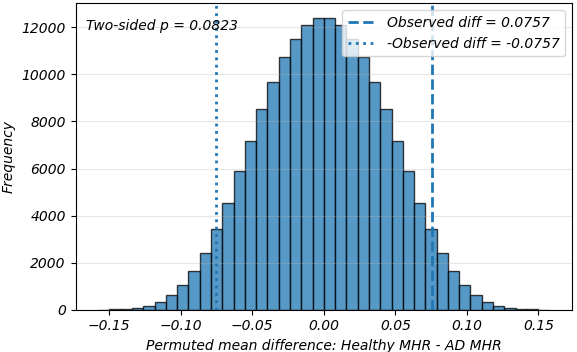}
        \caption{Permutation of MHR calculated using STAR-QSM}
        \label{fig:star_permutation_mhr}
    \end{subfigure}

    \caption{Permutation Test of MHR values of Healthy and AD groups using DECOMPOSE-QSM and STAR-QSM .}
    \label{fig:mhr_permutation}
\end{figure}

\begin{figure}[htbp]
    \centering

    \begin{subfigure}{0.49\columnwidth}
        \centering
        \includegraphics[width=\linewidth]{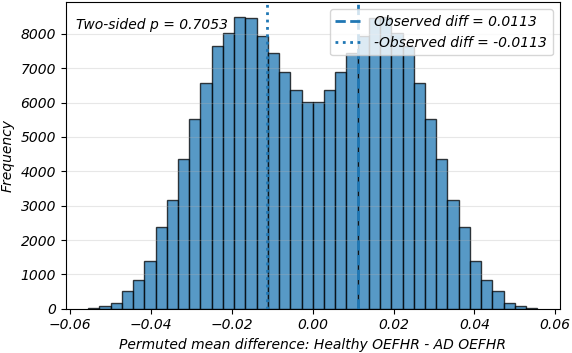}
        \caption{Permutation of OEFHR calculated using DECOMPOSE-QSM}
        \label{fig:decompose_permutation_oefhr}
    \end{subfigure}
    \hfill
    \begin{subfigure}{0.49\columnwidth}
        \centering
        \includegraphics[width=\linewidth]{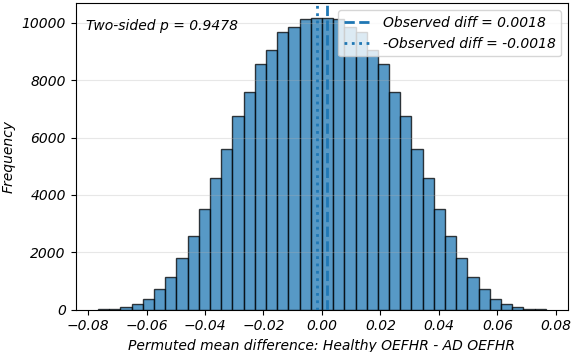}
        \caption{Permutation of OEFHR calculated using STAR-QSM}
        \label{fig:star_permutation_oefhr}
    \end{subfigure}

    \caption{Permutation Test of OEFHR values Healthy and AD groups using DECOMPOSE-QSM and STAR-QSM .}
    \label{fig:oefhr_permutation}
\end{figure}

\begin{table}[htbp]
\centering
\caption{Myelin Hemispheric Ratio (MHR)-based statistical comparison between Healthy and AD groups.}
\label{tab:hr_random_subset_stats}
\scriptsize
\resizebox{\textwidth}{!}{
\begin{tabular}{lcccccccccc}
\toprule
\textbf{Method} &
\makecell{\textbf{Healthy MHR}\\\textbf{Mean $\pm$ SD}} &
\makecell{\textbf{AD MHR}\\\textbf{Mean $\pm$ SD}} &
\makecell{\textbf{F-test}\\\textbf{p}} &
\makecell{\textbf{Welch}\\\textbf{p}} &
\makecell{\textbf{Student}\\\textbf{p}} &
\makecell{\textbf{Mann--Whitney}\\\textbf{p}} &
\makecell{\textbf{Permutation}\\\textbf{p}} &
\makecell{\textbf{ANOVA}\\\textbf{F}} &
\makecell{\textbf{ANOVA}\\\textbf{p}} &
\makecell{\textbf{Hedges'}\\\textbf{$g$}} \\
\midrule
DECOMPOSE-QSM
& $1.0500 \pm 0.0399$
& $1.0291 \pm 0.0401$
& $0.9911$
& $0.2586$
& $0.2586$
& $0.3847$
& $0.2630$
& $1.3607$
& $0.2586$
& $0.4996$
\\

STAR-QSM
& $1.1367 \pm 0.1050$
& $1.0611 \pm 0.0783$
& $0.3942$
& $0.0857$
& $0.0843$
& $0.0890$
& $0.0823$
& $3.3387$
& $0.0843$
& $0.7826$
\\
\bottomrule
\end{tabular}
}
\end{table}

\begin{table}[htbp]
\centering
\caption{Mixed $2 \times 2$ ANOVA on MHR with Group as the between-subject factor and Method as the within-subject factor.}
\label{tab:hr_mixed_anova_random_subset}
\scriptsize
\resizebox{0.66\columnwidth}{!}{
\begin{tabular}{lcccc}
\toprule
\textbf{Effect} &
\textbf{F} &
\textbf{df} &
\textbf{p-value} &
\textbf{Partial $\eta^2$} \\
\midrule
Group &
2.742 &
$(1,18)$ &
0.1151 &
0.132 \\
Method &
21.000 &
$(1,18)$ &
0.0002 &
0.538 \\
Group $\times$ Method &
4.477 &
$(1,18)$ &
0.0486 &
0.199 \\
\bottomrule
\end{tabular}
}
\end{table}

\begin{table}[htbp]
\centering
\caption{Oxygen Extraction Fraction Hemispheric Ratio (OEFHR)-based statistical comparison between Healthy and AD groups.}
\label{tab:oefhr_stats}
\scriptsize
\resizebox{\textwidth}{!}{
\begin{tabular}{lcccccccccc}
\toprule
\textbf{Method} &
\makecell{\textbf{Healthy OEFHR}\\\textbf{Mean $\pm$ SD}} &
\makecell{\textbf{AD OEFHR}\\\textbf{Mean $\pm$ SD}} &
\makecell{\textbf{F-test}\\\textbf{p}} &
\makecell{\textbf{Welch}\\\textbf{p}} &
\makecell{\textbf{Student}\\\textbf{p}} &
\makecell{\textbf{Mann--Whitney}\\\textbf{p}} &
\makecell{\textbf{Permutation}\\\textbf{p}} &
\makecell{\textbf{ANOVA}\\\textbf{F}} &
\makecell{\textbf{ANOVA}\\\textbf{p}} &
\makecell{\textbf{Hedges'}\\\textbf{$g$}} \\
\midrule

DECOMPOSE-QSM
& $1.0294 \pm 0.0601$
& $1.0181 \pm 0.0323$
& $0.0787$
& $0.6093$
& $0.6074$
& $0.7337$
& $0.7053$
& $0.2735$
& $0.6074$
& $0.2240$
\\

STAR-QSM
& $1.0359 \pm 0.0665$
& $1.0341 \pm 0.0423$
& $0.1936$
& $0.9435$
& $0.9434$
& $0.7913$
& $0.9478$
& $0.0052$
& $0.9434$
& $0.0309$
\\

\bottomrule
\end{tabular}
}
\end{table}

\begin{table}[htbp]
\centering
\caption{Mixed $2 \times 2$ ANOVA on OEFHR with Group as the between-subject factor and Method as the within-subject factor.}
\label{tab:oefhr_mixed_anova}
\scriptsize
\resizebox{0.66\columnwidth}{!}{
\begin{tabular}{lcccc}
\toprule
\textbf{Effect} &
\textbf{F} &
\textbf{df} &
\textbf{p-value} &
\textbf{Partial $\eta^2$} \\
\midrule
Group &
0.081 &
$(1,18)$ &
0.7790 &
0.004 \\
Method &
7.882 &
$(1,18)$ &
0.0116 &
0.305 \\
Group $\times$ Method &
1.385 &
$(1,18)$ &
0.2546 &
0.071 \\
\bottomrule
\end{tabular}
}
\end{table}

Both the results suggest that the hemispheric asymmetry is attenuated rather than reversed when affected by Alzheimer's disease. The attenuation in both MHR and OEFHR due to AD can be potential biomarkers of the disease itself. The attenuation may not be statistically significant due to the sample size being significantly small for statistical analysis.

\begin{figure}[htbp]
    \centering

    \begin{subfigure}{0.49\columnwidth}
        \centering
        \includegraphics[width=\linewidth]{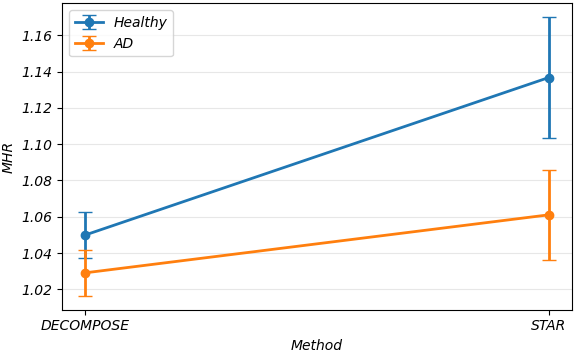}
        \caption{MHR}
        \label{fig:interaction_mhr}
    \end{subfigure}
    \hfill
    \begin{subfigure}{0.49\columnwidth}
        \centering
        \includegraphics[width=\linewidth]{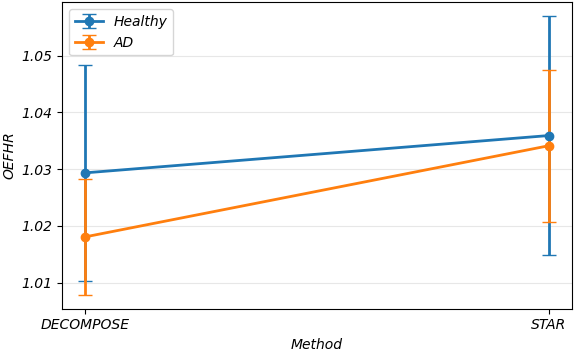}
        \caption{OEFHR}
        \label{fig:interaction_oefhr}
    \end{subfigure}

    \caption{Group $\times$ Method interaction plots.}
    \label{fig:interaction_plots}
\end{figure}

The Box Plots of Fig-~\ref{fig:mhr_boxplots} and Fig-~\ref{fig:oefhr_boxplots} suggest that there is a methodological difference in value when we are calculating OEFHR and MHR for both healthy controls and AD patients. The claim is strengthened by the plots in Fig-~\ref{fig:interaction_plots} where we can see that for MHR, the gap between AD and Healthy is more in STAR-QSM compared to DECOMPOSE-QSM and the MHR values are elevated in STAR-QSM. On the other side, for OEFHR, the gap between AD and Healthy is more in DECOMPOSE-QSM compared to STAR-QSM and just like MHR, the OEFHR values are elevated in STAR-QSM. Table-~\ref{tab:hr_mixed_anova_random_subset} and Table-~\ref{tab:oefhr_mixed_anova} likewise suggest that OEFHR and MHR both are strongly influenced by the QSM method used. However, for MHR the methodological effect differs between Healthy and AD whereas for OEFHR this difference in methodological effect is not statistically significant.

The white-matter asymmetry in healthy controls is also mentioned and discussed in several research works \cite{Kanakaraj2025LifespanAsymmetryTracts, Parekh2023WhiteMatterLateralizationSPD, Wan2024MicrostructuralAsymmetry, Korbmacher2024BrainAsymmetries, Toga2003, Kong2022}. This phenomenon is also verified by our findings where we found that the Myelin Hemispheric Ratio is more than one for all healthy controls as understood from the box-plots of Fig-~\ref{fig:mhr_boxplots}. Prominent research also suggest that this hemispheric asymmetry is altered in AD patients \cite{Yang2017, Chandio2026AlongTractAsymmetryAD}. This trend is present in our findings as we can see  from the box-plots of Fig-~\ref{fig:mhr_boxplots}, the mean MHR in AD is lower than that of healthy controls. More importantly, some AD subjects have an MHR value of less than one. Though the results are not statistically significant, they definitely suggest that AD alters or possibly reduces the hemispheric asymmetry which is found in a healthy brain. Turning to OEFHR, the analysis both reproduces and extends this picture which can be understood from the box plots of Fig-~\ref{fig:oefhr_boxplots}. The right-dominant asymmetry seen in the myelin-based MHR is recovered on OEFHR, being reliable in controls and attenuated in AD, so the loss of lateralisation in disease is not confined to a single tissue property. Lowered neural activity leads to reduced oxygen demand \cite{Jiang2020ADVascular, Thomas2017MCIMetabolism, Hosseini2023QSMOEFADSpectrum}. DECOMPOSE-QSM leads to better estimation of the ratios as source mixing may lead to less accurate calculations due to paramagnetic venous signals being cancelled partially by co-localised diamagnetic myelin \cite{AHMED2023120357}. Reduced OEFHR which suggests attenuated asymmetry is compatible with a metabolic component of AD pathology, since oxidative metabolism sustains oligodendrocyte function and myelin maintenance \cite{Yan2025OEFWhiteMatterDysfunction, Misra2026OEFPathologicalBiomarkers}.

The purpose of this study was to understand the difference of hemispheric asymmetry in healthy controls and AD patients that maybe potentially utilized to understand the disease pathology of AD. The study aims to understand the differences in the level of neuro-degeneration caused by Alzheimer's disease in both the hemispheres. Even though the study showed differences in hemispheric asymmetry in healthy controls and AD patients, the difference is not statistically significant. However, the sample size being very small maybe one of the main reasons of lack of statistical significance. Future directions of this study include calculating the MHR and OEFHR of AD patients and healthy controls by taking a bigger sample size which will lead to finding out whether the statistical insignificance is due to the sample size or the differences in hemispheric asymmetry, that is, the attenuated values of MHR and OEFHR in AD are not statistically significant at all. Besides, the study aims to show the differences in numerical values of MHR and OEFHR for bulk QSM and source-separated QSM in case of healthy controls and AD patients. This helps to understand how source separation at sub-voxel level leads to different estimation of quantitative markers compared to bulk QSM.  

\section*{Acknowledgment}

The authors thank UC Berkeley and UC San Francisco for the dataset used in this study. The authors extend their heartfelt gratitude to Professor Dr. Chunlei Liu of UC Berkeley for his valuable insights and advice throughout the duration of the study. The authors also express their gratitude to BUET for providing the workstation computer used in the study.

\bibliographystyle{IEEEtran}
\bibliography{ref}

\end{document}